\documentclass[aps,prd,showpacs,floatfix,twocolumn]{revtex4}

\usepackage[dvips]{color}
\usepackage[dvips]{epsfig}
\usepackage{amsmath}
\usepackage{amssymb}

\begin{document}

\title{Analysis of four-wave mixing of high-power lasers for the detection of elastic photon-photon scattering}

\author{J.\ Lundin, M.\ Marklund, E.\ Lundstr\"om\footnote{Currently at: 
Department of Physics, Stockholm University, SE--106 91,
Sweden}, G.\ Brodin}
\affiliation{Department of Physics, Ume{\aa} University, SE--901 87 Ume{\aa},
Sweden}




\author{J.\ Collier, R.\ Bingham, J.T.\ Mendon\c{c}a, P.\ Norreys}
\affiliation{Rutherford Appleton Laboratory, Chilton, Didcot
OX11 0QX, Oxfordshire, UK}




\date{\today}

\begin{abstract}
We derive expressions for the coupling coefficients for electromagnetic four--wave mixing in the non--linear quantum vacuum. An experimental setup for detection of elastic photon--photon scattering is suggested, where three incoming laser pulses collide and generate a fourth wave with a new frequency and direction of propagation. An expression for the number of scattered photons is derived and, using beam parameters for the Astra Gemini system at the Rutherford Appleton Laboratory, it is found that the signal can reach detectable levels. Problems with shot--to--shot reproducibility are reviewed, and the magnitude of the noise arising from competing scattering processes is estimated. It is found that detection of elastic photon--photon scattering may for the first time be achieved.
\end{abstract}
\maketitle

\section{Introduction}
The linearity of Maxwell's vacuum equations does not allow electromagnetic waves propagating in vacuum to interact, and therefore photon--photon scattering is forbidden in classical physics. However, according to the theory of quantum electrodynamics (QED), the quantum vacuum possesses nonlinear properties. Photon--photon scattering may therefore occur, owing to the interaction with virtual electron--positron pairs. It is thought that QED non--linear vacuum effects may be of importance in the neighborhood of strongly magnetized astrophysical systems, such as pulsars \cite{Beskin, Curtis} and magnetar environments \cite{Harding, Kouveliotou}. With the rapidly growing power of present day laser systems \cite{Tajima}, we are at the brink of making the non--linear properties of the quantum vacuum directly accessible for observations \cite{Ringwald, Lundstrom}. Effects such as electron--positron pair creation and elastic photon--photon scattering may even play an important role in future laser--plasma experiments \cite{Marklund}. One such example would be laser self focusing \cite{Bulanov, Shorokhov}, where laser pulse compression gives rise to field strengths close to the critical field strength, $E_{\text{crit}}\approx 10^{18}\,\mathrm{V\,cm^{-1}}$. 

Direct observation of elastic photon--photon scattering among real photons would, because of its fundamental importance to QED, be of great scientific importance. Throughout the last decades, several suggestions on how to detect elastic photon--photon scattering have been made. For example, using harmonic generation in an inhomogeneous magnetic field \cite{Recent3}, using resonant interaction between eigenmodes in microwave cavities \cite{Brodin-Marklund-Stenflo,EBMS}, using ultra--intense fields occurring in plasma channels  \cite{Shen-Yu-Wang}, as well as many others, see e.g. Refs. \cite{Older2,Recent4,Recent6}. However, so far no suggestions have led to actual detection of photon--photon scattering among real photons. Related processes, but physically different from elastic photon--photon scattering, is photon splitting \cite{Adler} (see also \cite{Gubbe}) and Delbr\"uck scattering \cite{Jarlskog}, of which the latter has been detected using high--$Z$ atomic targets \cite{Jarlskog}.

The present paper is an expansion of Ref.\ \cite{Lundstrom}, where the possibility to detect photon--photon scattering in vacuum using four--wave mixing is investigated and a concrete experimental configuration is suggested where three colliding laser pulses stimulate emission in a fourth direction with a new frequency. Both theory, argumentation and the results are in the present paper presented in more depth than before, but the paper also features new aspects, e.g. the analysis of the expected shot--to--shot reproducibility in the suggested experiment, the analysis of alternative experimental configurations and an extended discussion regarding noise sources.

Using four--wave mixing to stimulate photon--photon scattering has the advantage of not being limited by the low scattering cross section, $\sigma_{\gamma\gamma}\approx0.7\times10^{-65}\ \mathrm{cm^{\!2}}$ \cite{Berestetskii}, in the optical range. The idea of QED four--wave mixing was first studied by \cite{Adler70}. Further theoretical \cite{Recent,Moulin-Bernard,DiPiazza} and experimental \cite{Bernard-exp} studies have been performed since then. The main obstruction for successful detection of scattering events has been the lack of sufficiently powerful laser technology. Since present petawatt laser systems only operate at a single frequency (with harmonics available through the use of frequency doubling crystals), the configuration of beams and the geometry of the setup also becomes important. We show that any two--dimensional (2D) setup is unlikely to produce a signal distinguishable from noise, for presently available power levels. This is in contrast to the 3D setup presented in \cite{Lundstrom} and in this paper. Here we have calculated the coupling coefficient for four--wave mixing as a function of incident angles and laser polarization. This is done both for the general 2D setup and a specific 3D setup. We have also determined an easy--to--use expression for the number of scattered photons with the 3D configuration, given specific beam data. The beam data parameters are chosen to fit the high--repetition rate Astra Gemini system (operational 2007) located at the Central Laser Facility (Rutherford Appleton Laboratory), giving an estimated number of 0.07 scattered photons per shot. Furthermore, a unique fingerprint of the QED process can be obtained through the polarization dependence of the number of scattered photons.

In order to get a statistically sufficient number of scattered photons, a good shot--to--shot reproducibility is required. The shot--to--shot reproducibility of the Astra Gemini system is analysed, and it is found that energy fluctuations will be small on a shot--to--shot basis. While vibrations could pose a problem, they are expected to be small due to the inherent structure of the Astra Gemini system and the experimental setup. Further ways of improving the vibration stability, if necessary, is also suggested.

Due to a non--perfect vacuum in the interaction chamber, competing scattering processes such as Compton scattering and in principle collective plasma four--wave mixing will be present. An analysis, with the help of computer simulations, has been carried out in order to give a rough estimate of the signal to noise ratio. It is found that plasma cavitation caused by the strong laser pulses suppress the competing effects rather effectively close to the interaction region (see e.g. Ref. \cite{Guzdar}). Specifically, in our suggested experimental setup, this reduces the noise to a level approximately three orders of magnitude lower than the QED signal. Thus, we conclude that detection of photon--photon scattering will in principle be possible with the Astra Gemini system. 

\section{Theory}
It is possible to effectively relate the properties of the QED-vacuum to those of a medium in ordinary classical electrodynamics. By integrating out all high energy degrees of freedom, an effective field theory containing only the electromagnetic fields can be formulated in terms of the Heisenberg--Euler Lagrangian \cite{Heisenberg-Euler,Schwinger}, valid for field strengths below the QED critical field $10^{16}\,\mathrm{V\,cm^{-1}}$ and for wavelengths larger than the Compton wavelength $10^{-10}\,\mathrm{cm}$ \cite{general}. The Heisenberg--Euler Lagrangian is given by
\begin{eqnarray}
  &&\!\!\!\!\!\!	{\cal L}={\cal L}_{0}+\delta{\cal L}
  \nonumber \\ && \!\!\!\!
   =\frac{1}{8\pi}\left(E^{2}\!-B^{2}\right)+\frac{\xi}{8\pi}\left[\left(E^{2}\!-B^{2}\right)^{2}+7\left(\textbf{E}\cdot\textbf{B}\right)^{2}\right] ,
\end{eqnarray} 
where $\xi=\hbar e^{4}/45\pi m^{4}c^{7}$, $\hbar = h/2\pi$, $h$ is Planck's constant, 
$e$ is the magnitude of the electron charge, $m$ is the electron mass, and $c$
is the speed of light in vacuum. ${\cal L}_{0}$ is the Lagrangian for classical electrodynamics, while $\delta{\cal L}$ represents the lower order non--linear QED--correction. 

On identifying an effective polarization and magnetization of the vacuum as
\begin{equation}\label{eq:polarization}
\textbf{P}=\frac{\xi}{4\pi}\left[2\left(E^{2}\!-B^{2}\right)\textbf{E}+7\left(\textbf{E}\cdot\textbf{B}\right)\textbf{B}\right] 
\end{equation}
and
\begin{equation}\label{eq:magnetization}
\textbf{M}=\frac{\xi}{4\pi}\left[-2\left(E^{2}\!-B^{2}\right)\textbf{B}+7\left(\textbf{E}\cdot\textbf{B}\right)\textbf{E}\right],
\end{equation}
the equations of motion following from the Heisenberg--Euler Lagrangian can be written
\begin{eqnarray}
	\nabla\cdot\textbf{E}=-4\pi\nabla\cdot\textbf{P},
\end{eqnarray}
\begin{eqnarray}
	\nabla\times\textbf{E}+\frac{1}{c}\frac{\partial\textbf{B}}{\partial t}=0,
\end{eqnarray}
\begin{eqnarray}
	\nabla\cdot\textbf{B}=0,
\end{eqnarray}
\begin{eqnarray}
	\nabla\times\textbf{B}-\frac{1}{c}\frac{\partial\textbf{E}}{\partial t}=4\pi\frac{1}{c}\frac{\partial\textbf{P}}{\partial t}+4\pi\nabla\times\textbf{M},
\end{eqnarray}
that is Maxwell's equations in the presence of a medium with polarization $\textbf{P}$ and magnetization $\textbf{M}$. From those we can derive the effective wave equation
\begin{equation}\label{eq:wave-equation}
	\Box\textbf{E} 
	= 4\pi c^{2}\!\left[\nabla \left(\nabla\cdot\textbf{P}\right)-\frac{1}{c}\frac{\partial}{\partial t}\left(\frac{1}{c}\frac{\partial\textbf{P}}{\partial t}+\nabla\times\textbf{M}\right)\right],
\end{equation}
where $\Box = \partial_t^{2} - c^2\nabla^{2}$.

Next we consider three plane waves, representing the incoming laser pulses, with amplitudes allowed to have a weak space time dependence due to interactions,
\begin{equation}
\textbf{E}_{j}\!\left(\textbf{r},t\right) = \tfrac{1}{2}\left(\tilde{\textbf{E}}_{j}\!\left(\textbf{r},t\right)e^{i\varphi_j(\mathbf{r},t)}+\tilde{\textbf{E}}^{*}_{j}\!\left(\textbf{r},t\right)e^{-i\varphi_j(\mathbf{r},t)}\right) , 
\end{equation}
where $\varphi_j = \textbf{k}_{j}\cdot\textbf{r}-\omega_{j}t $ and $j=1,2,3$. 
Due to the cubic non--linearity of (\ref{eq:polarization}) and (\ref{eq:magnetization}), 
we expect generation of a fourth wave with $(\omega_{4},\textbf{k}_{4})$ if the incoming wave vectors satisfy the matching conditions
\begin{equation}\label{eq:wavevector}
	\textbf{k}_{1}+\textbf{k}_{2}=\textbf{k}_{3}+\textbf{k}_{4} 
\end{equation}
and 
\begin{equation}\label{eq:frequency}
	\omega_{1}+\omega_{2}=\omega_{3}+\omega_{4},
\end{equation}
for some $\textbf{k}_{4}$ and $\omega_{4}\!=\!ck_{4}$. In practice the interaction pulses will have a finite spectral width, $\Delta \textbf{k}_{j}$ and equations (\ref{eq:wavevector}) and (\ref{eq:frequency}) refers to the central peaks of the spectra. However, we assume that the spectral width is small ($\left|\Delta \textbf{k}_{j}\right|\ll\left|\textbf{k}_{j}\right|$), such that it can be incorporated into slowly varying amplitudes, $\left|\partial E_{j}/\partial t\right| \ll \omega_{j}\left|E_{j}\right|,\left|\nabla E_{j}\right| \ll \left|k_{j}\right|\left|E_{j}\right|$ and the derivatives can be taken to act only on the harmonic parts. The effect of a finite spectral width is to give a finite interaction time. Note that for a sufficiently broad spectra, more general methods must be used, see e.g. \cite{tito}. From (\ref{eq:wavevector}) and (\ref{eq:frequency}) we find that only the resonant terms including the factors $\tilde{E}_{1}\tilde{E}_{2}\tilde{E}^{*}_{3}$ and $\tilde{E}^{*}_{1}\tilde{E}^{*}_{2}\tilde{E}_{3}$ are of importance, since all others will average to zero over a short spacetime interval due to rapid oscillations.

Hence, on neglecting the non-resonant terms, the wave equation (\ref{eq:wave-equation}) for the generated field $\textbf{E}_{g}$ takes the form

\begin{equation}\label{eq:wave-equation2}
\Box\bar{\textbf{E}}_{g}\!\left(\textbf{r},t\right)=4\xi\omega^{2}_{4}\textbf{G}\tilde{E}_{1}\tilde{E}_{2}\tilde{E}^{*}_{3}e^{i\left(\textbf{k}_{4}\cdot\textbf{r}-\omega_{4}t\right)}
\end{equation}
where
\begin{eqnarray}
\textbf{E}_{g}\!\left(\textbf{r},t\right)=\frac{1}{2}\left(\bar{\textbf{E}}_{g}\!\left(\textbf{r},t\right)+\bar{\textbf{E}}^{*}_{g}\!\left(\textbf{r},t\right)\right)
\end{eqnarray}
and $\textbf{G}$ is a geometric factor which only depends on the directions of the wave vectors and polarization vectors, and can be solved for from (\ref{eq:wave-equation}). The driving of the initial waves can be neglected due to the weakness of the generated field, and hence the  energy in each pulse is constant during the interaction.

The radiation zone solution to the wave equation (\ref{eq:wave-equation2}) takes the form of a spherical outgoing wave multiplied by a direction dependent factor peaked in the resonant $\hat{\textbf{k}}_{4}$-direction,
\begin{eqnarray}\label{eq:wave-equation2solution}
  && \bar{\textbf{E}}_{g}(\textbf{r},t) = 
  \frac{\xi k^{2}_{4}}{\pi r}\textbf{G}e^{i\left(k_{4}r-\omega_{4}t\right)}
  \nonumber \\ &&\quad \times \int_{V'}\left.\left(\tilde{E}_{1}\tilde{E}_{2}\tilde{E}^{*}_{3}\right)\right|_{t_{R}}e^{i{k}_{4}\left(\hat{\textbf{k}}_{4}-\hat{\textbf{r}}\right)\cdot\textbf{r}'}dV',
\end{eqnarray}
where $V'$ is the interaction region, and the amplitudes are to be evaluated at the retarded time $t_{R}=t-\left|\textbf{r}-\textbf{r}'\right|/c$.

\begin{figure}[ht]
\includegraphics[width=0.9\columnwidth]{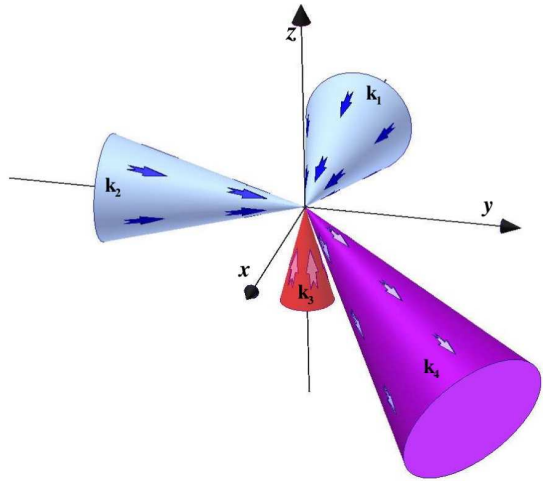}
\caption{Configuration of the incoming laser beams and the direction of the scattered wave for the wave vectors defined in (\ref{eq:specific-wavevectors}), which satisfies the matching conditions (\ref{eq:wavevector}) and 
(\ref{eq:frequency}).}
\label{fig:3d}
\end{figure}

Equation (\ref{eq:wave-equation2solution}) holds for any set of wave vectors satisfying (\ref{eq:wavevector}) and (\ref{eq:frequency}), but for real experiments only a few configurations are of interest. In practice, the highest power laser systems operates at a rather well--defined 
frequency, and by using frequency--doubling crystals we also have access to second harmonics, although
with some power loss. As will be argued below, these restrictions in incoming frequencies make a 2D setup less competitive than a 3D setup. The 3D configuration achievable from a single laser through frequency doubling which is best suited for an experiment (allowed by the matching conditions (\ref{eq:wavevector}) and (\ref{eq:frequency})) is given by the wavevectors
\begin{eqnarray}\label{eq:specific-wavevectors}
	\left.\begin{array}{l}\textbf{k}_{1}=k\hat{\textbf{x}}\\ \textbf{k}_{2}=k\hat{\textbf{y}}\\ \textbf{k}_{3}=\frac{k}{2}\hat{\textbf{z}}\\ \textbf{k}_{4}=k\hat{\textbf{x}}+k\hat{\textbf{y}}-\frac{1}{2}k\hat{\textbf{z}}\end{array}\right\},
\end{eqnarray}
and is illustrated in Fig.\ \ref{fig:3d}.

The main advantages of using a configuration defined by (\ref{eq:specific-wavevectors}) are that the wavelength and the direction of the generated wave are well separated from that of the incoming laser beams. This is highly desirable since optical components will always scatter and reflect some light, making it difficult to distinguish QED scattered photons of wavelengths close to that of the incoming laser beams. Furthermore, the geometry of the setup is not hindered by the fact that the focused laser beams in reality are cone--shaped. Also, the polarizations can be chosen arbitrary since it is not necessary to prevent any counter propagating laser beam from entering back into the laser.

In order to calculate the geometric factor $\textbf{G}$ we define $\gamma_{j}$ as the angle between the \textit{z}--axis and the polarization vector of the $E_{j}$--field, in such way that it is positive in clockwise direction when looking in the $\textbf{k}_{j}$--direction. $\beta_{3}$ is defined in the same manner but with respect to the \textit{x}--axis. 
In the case of configurations confined to a plane, the wave vectors are given by
\begin{eqnarray}
	\textbf{k}_{j}=k_{j}\cos\phi_{j}\hat{\textbf{x}}+k_{j}\sin\phi_{j}\hat{\textbf{y}},
\end{eqnarray}
where $\phi_{j}$ is the angle from the \textit{x}-axis to $\textbf{k}_{j}$. Using these
definitions,  
the general 2D and the specific 3D geometric factors are given by (\ref{eq:2d}) and (\ref{eq:3d}) \cite{Erik}, respectively, in the Appendix. (\ref{eq:2d}) is valid for any set of wave vectors confined to a plane as long as (\ref{eq:wavevector}) and (\ref{eq:frequency}) are satisfied, while (\ref{eq:3d}) is valid for the specific set of wave vectors given by (\ref{eq:specific-wavevectors}). Note that the 2D case describes the general four--wave process in the center--of--mass frame of two incoming photons, with wave vectors $\textbf{k}_{1}'$ and $\textbf{k}_{2}'=-\textbf{k}_{1}'$. Thus, the general 3D geometric factor $\textbf{G}$ can be found from the 2D one by applying Lorentz transformations.

The number of scattered photons can be estimated from modeling the laser pulses as rectangular prisms with length $L$ and quadratic cross section $b^{2}$, inside which the field amplitudes are constant. Although not a completely accurate model of the pulse shape, it has the advantage of making the calculations transparent, and it will give a reasonable estimate of the number of generated photons.

In the 3D case the interaction region will take the shape of a cube with side $b$, existing during a time $L/c$. By carrying out the integration in (\ref{eq:wave-equation2solution}), the generated electric field can now be solved for, from which the generated intensity and total power can be calculated. After multiplying the generated power by the interaction time and dividing by the photon energy $\hbar\omega_{4}$, the estimated number of scattered photons per shot is found to be
\begin{equation}\label{eq:number}
	N_{3d}=1.31\eta^{2} G_{3d}^{2}\left(\frac{1\,{\rm \mu m}}{\lambda_{4}}\right)^{3}\left(\frac{L}{1\,{\rm\mu m}}\right)\left(\frac{P_1P_2P_3}{1\,{\rm PW}^3}\right)
\end{equation}
where $G_{3d}$ is the geometric factor $|\textbf{G}_{3d}|$ which is given by (\ref{eq:3d}), $P_{j}$ is the power of the incoming pulses, $\lambda_{4}$ the generated wavelength, and with $\eta^{2}$ given by 
\begin{eqnarray}
  && \eta^{2} \equiv 
  \int^{2\pi}_{0}\int^{\pi}_{0}\frac{I_{g}\left(r,\theta,\phi\right)}{I_{g,res} 
  \left(r\right)}\sin\theta~d\theta ~d\phi
  \nonumber \\ && \!\!\!\! = \frac{64}{k^{6}_{4}b^{6}}\int^{2\pi}_{0}\int^{\pi}_{0} 
  \frac{\sin^{2}\left[k_{4}bf(\theta,\phi)/2\right]}{f^2(\theta,\phi)}\frac{\sin^{2}\left[k_{4}bg(\theta,\phi)/2\right]}{g^2(\theta,\phi)}\nonumber \\ &&\!\!\!\!\times\frac{\sin^{2}\left[k_{4}\frac{b}{2}\cos\theta\right]}{\cos^{2}\theta}\sin\theta ~d\theta ~d\phi
  \nonumber \\ && \!\!\!\!
\approx 0.025\left(\frac{\lambda_{4}}{0.267\,{\rm \mu m}}\right)^2\left(\frac{1.6\,{\rm \mu m}}{b}\right)^2,
\end{eqnarray}
where $I_{g}$ is the generated intensity, $I_{g,res}$ its maximum value 
in the resonant direction, $f(\theta,\phi) = 1-\cos\phi\sin\theta$, and 
$g(\theta,\phi) = \sin\phi\sin\theta$. The approximate equality of $\eta^{2}$ is accurate within $7\%$ for $b/\lambda_{4}<30$, i.e. $b/\lambda_{3}<10$ where $\lambda_{3}$ is the fundamental wavelength.
$\eta$ corresponds to roughly half the angular width of the central interference peak of the generated intensity.

The scaling of the number of generated photons with respect to $G_{3d}$ is decoupled from the pulse model. Hence it is possible to test the theoretical predictions by varying the polarizations of the incoming pulses. However, changing the linear polarization of one of the laser beams could create a false periodic signal, as the polarization affects the propagation direction of the competing Compton scattered photons. A low noise contribution is therefore a necessity. The upper panel of Fig.\ \ref{fig:nr-of-photons} shows how the scattering number depends on the polarization angle of wave three, with the polarization of the other two waves fixed at $\gamma_{1}=0$ and $\gamma_{2}=\pi/2$. An optimal choice of polarization angles is given by $\gamma_{1}=0$ and $\gamma_{2}=\beta_{3}=\pi/2$, for which $G^{2}_{3d}$=\ 0.77.

Should the polarization dependence from experiments deviate from that predicted by theory, it would be interesting to know whether this deviation also suggests deviation from Lorentz invariance. Lagrangians of Born--Infeld type are Lorentz invariant and can give us some insight in how to interpret a deviation. Nonlinear electrodynamics was suggested by Born \& Infeld \cite{born-infeld}, and later reformulated by Dirac \cite{dirac}, already in 1934, as a means for obtaining finite field energies around singular charges. In essence, the Born-Infeld Lagrangian builds on the Lorentz invariants constructed from the Maxwell field tensor, and can be written
\cite{dirac}
\begin{equation}\label{eq:born-infeld}
    {\cal L} = \kappa^{2}\left[1-\sqrt{1+\frac{1}{2\kappa^{2}}\left(E^{2}\!-B^{2}\right)-\frac{1}{16\kappa^{4}}\left(\textbf{E}\cdot\textbf{B}\right)^{2}}\right],
\end{equation}
where 
$\kappa$ is the relevant coupling constant for the modified electrodynamics. The Born-Infeld form of Lagrangian (\ref{eq:born-infeld}) also occurs as the effective field limit of quantized strings \cite{fradkin-tsetylin} (see Ref.\ \cite{tsetylin} for a review). In such effective Lagrangians the field strength coefficients contain the string tension. We note that (\ref{eq:born-infeld}) gives no birefringence in static fields, a unique property of the Born--Infeld Lagrangian \cite{Kerner}. In general one speaks of Born-Infeld type Lagrangians when we have Lagrangian built from the Lorentz field invariants, and the coefficients are kept arbitrary, see (\ref{eq:Born-Infeld-Lagrangian}). A Lagrangian of this form affects the Geometric factor, and thus the polarization dependence of the signal as described in Appendix A. Any detected signal that can be fit to follow the polarization dependence of the geometric factors for some $X$ and $Y$ is thus consistent with Lorentz invariance.

\section{Experimental parameters}
\begin{figure}[ht]
\includegraphics[width=0.75\columnwidth]{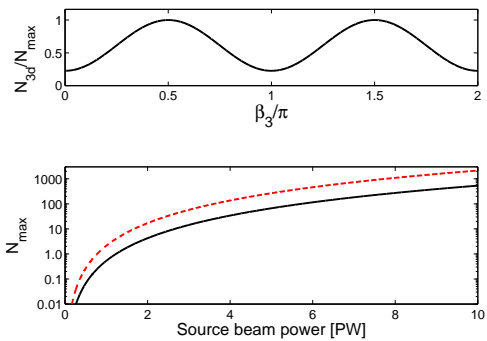}
\caption{The upper panel shows the number of scattered photons $N_{3d}$ per shot, normalized by the number of photons $N_{\text{max}}$ per shot for an optimal choice of polarization, as a function of polarization angle $\beta_{3}$ of the wave with direction $\hat{\mathbf{k}}_3$. The lower panel shows $N_{\text{max}}$ predicted by (\ref{eq:number}) when increasing the laser power while keeping the beam width constant at $b=1.6\ \mathrm{\mu m}$, for a system where two source beams are used and one of the beams is split into two (solid line) and when three source beams are used, hence no beam splitting is required (dashed line).}
\label{fig:nr-of-photons}
\end{figure}

The Astra Gemini Laser will generate two independently configurable 0.5 PW laser beams of wavelength $800$ nm. Each pulse will contain a total energy of 15 J, and focused intensities over $10^{22}\ \mathrm{Wcm^{\!-2}}$ will be reached. The repetition rate is expected to be three shots per minute. Using these values, the spatial dimensions of the pulse model are taken as $b=1.6\ \mathrm{\mu m}$ and $L=10\ \mathrm{\mu m}$, which gives $\eta^{2}=0.025$. The 3D configuration described by (\ref{eq:specific-wavevectors}) can be achieved if one of the laser beams is frequency doubled and split into two beams. The estimated loss of power when frequency doubling is approximately 60\%, and hence the power of the incoming beams are given by $P_{1}=P_{2}=0.1$ PW, $P_{3}=0.5$ PW. The number of QED scattered photons, using optimal polarizations, would then be $N_{3d} =0.07$ and their wavelengths will be centered around 267 nm. The lower panel of Fig.\ \ref{fig:nr-of-photons} and Table \ref{tab:nr-of-photons} shows how the scattering number increases with increasing laser power, when the focused beam width is kept constant. 

We note that in order to get a statistically sufficient number of scattered photons, the conditions must be reproduced from shot to shot. The Astra Gemini system is designed to operate in saturation, so energy fluctuations are likely to be less than 5\% on a shot to shot basis. As for the pulse length, it primarily derives from static and fixed effects and fluctuations are expected to be less than 5\%, giving a maximum total fluctuation in power of about 10\%. The most important effect is that of spot and pointing stability. Through the use of adaptive optics, there will be active control of the spot size, shape and to some limited extent position. Controlling the relative pointing of the three beams after they have been split and frequency converted is a challenge. This will be done at large aperture so there is an inherent sensitivity to small pointing variations arising from vibration. Vibration could pose a problem, but the experiment should be designed in such a way that any given mode of vibration produces a similar far field displacement, i.e. all beams move in the same direction vertically or horizontally against a stimulus. Furthermore, the Astra Gemini system has been designed on very thick concrete slabs so vibration after this split is likely to be a minimum problem. Should this prove not to be the case, then commercially available stabilising systems would be used to stabilise the pointing of the beams. Finally, all of these parameters will be measured on each shot in order to sort the data accordingly. Below we will show that already the scattering number $N_{3d}=0.07$ for the AG system will exceed the estimated noise level.
 
\begin{table}
\caption{The table shows more precise values for the number of scattered photons in Fig.\ 2, for some different laser powers. $N_a$ is the number of scattered photons for a system where two source beams are used and $N_b$ is the number when three source beams are used. 0.066 is the expected number of scattered photons per shot with the Astra Gemini system.}
\begin{ruledtabular}
		\begin{tabular}{crr}
			Source beam power [PW] & $N_{a}$ & $N_{b}$\\[1mm]
			\hline
			0.5 & 0.066 & 0.27\\
			2.5 & 8.3 & 33\\
			5 & 66 & 266\\
			10 & $5.3\times 10^{2}$ & $2.1\times 10^{3}$\\
			25 & $8.3\times 10^{3}$ & $3.3\times 10^{4}$\\
			50 & $6.6\times 10^{4}$ & $2.7\times 10^{5}$\\
		\end{tabular}
\end{ruledtabular}
\label{tab:nr-of-photons}
\end{table}

\section{Alternative setups}
For comparison, let us consider a 2D configuration where two parallel beams collide head on with a third beam, all of them with the same wavelength. The interaction region will then take the shape of a rectangular prism of width $b$ and, with respect to the retarded time, length $L/2$, and exist for some time interval $L/c$. Since the pulse length often greatly exceeds the pulse width, this configuration gives an optimal interaction region and also generates the most scattered photons, $N=2.4$, for an optimal choice of polarization angles ($\gamma_{1}=\gamma_{3}=0$ and $\gamma_{2}=\pi/2$). However, the scattered photons are emitted along the beam axis and have the same wavelength as that of the incoming beams, making detection impossible in practice. The directions of the incoming and scattered waves can in principle be separated if their frequencies are shifted. The large cone angle of the laser beams and the need of having a well separated scattered frequency however requires a large deviation from the original configuration, thus destroying its nice features and consequently bringing down the scattering number. This also requires a greater change in frequency of the incoming beams, which may cause a problem with finding suitable lasers.

\begin{figure}[ht]
\includegraphics[width=0.45\columnwidth]{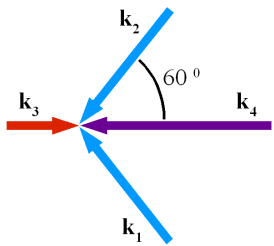}
\caption{An alternative experimental configuration for detection of photon-photon scattering through four-wave coupling.}
\label{fig:alternative}
\end{figure}

There is, however, a possible 2D configuration where only the fundamental frequency and its second harmonic is used, with the wavevectors
\begin{eqnarray}\label{eq:specific-wavevectors3}	\left.\begin{array}{l}\textbf{k}_{1}=-\frac{k}{2}\hat{\textbf{x}}+\frac{\sqrt{3}}{2}k\hat{\textbf{y}}\\ \textbf{k}_{2}=-\frac{k}{2}\hat{\textbf{x}}-\frac{\sqrt{3}}{2}k\hat{\textbf{y}}\\ \textbf{k}_{3}=\frac{k}{2}\hat{\textbf{x}}\\ \textbf{k}_{4}=-\frac{3}{2}k\hat{\textbf{x}}\end{array}\right\},
\end{eqnarray}
as illustrated in Fig.\ \ref{fig:alternative}. The geometric factor for this configuration becomes approximately $G^{2}_{2d}=5.5$, for an optimal choice of polarizations angles ($\gamma_1=\gamma_2=\gamma_3=3\pi/4$). Although this looks superior to the 3D setup, there are some practical problems. To extract the QED photons in this 2D setup, a beam splitter needs to be introduced in the path of $\hat{\textbf{k}}_3$, before the interaction region, bringing down the scattering number with a factor 0.25. Also, the interaction region in this case becomes somewhat smaller than a cubic one, further reducing the scattering number. Consequently, the estimated detectable signal will be of the same order as that of the 3D setup. The polarization dependence of the signal with this configuration is, however, much more pronounced than for the 3D one and it nearly vanishes at minimum. This setup could be an alternative to the 3D configuration presented above, but a major remaining problem is to ensure that none of the higher harmonic photons of the fundamental laser beam is reflected by optics back to the detector, drowning the signal in noise. 

There are other possible experimental configurations, both 2D and 3D, that we have not investigated in this paper. It is, for instance, possible to use third harmonics or higher, to construct setups with the scattering direction distinct from the incoming photon directions. However, this requires additional frequency doubling crystals reducing the power, and we expect these configurations to be less competitive. It is also possible to use configurations consisting of two or more lasers of different frequencies to generate QED scattered photons with wavelengths well separated from the harmonics of the incident beams, as used in the 3D configuration in Ref.\ \cite{Bernard-exp}. Since present laser systems have sufficient power only at single frequencies (with harmonics available using frequency--doubling crystals) we expect these configurations to be less competitive. Hence, the 3D configuration, where just the second harmonic is needed and the scattering direction is distinct from the directions of the incoming photons, seems to be the optimal setup from an experimental point of view.

\section{Noise sources}
Since a perfect vacuum in the interaction chamber is in practice impossible to achieve, there will be competing scattering processes present. The effect of these processes will rather effectively be suppressed due to plasma cavitation caused by the strong laser pulses. Below, with the help of computer simulations, we give a rough estimate of the magnitude of the competing scattering events resulting from Compton scattering and collective plasma four--wave mixing. This is in order to determine whether it is theoretically possible to distinguish the QED scattered photons from noise.

\subsection{Ponderomotive force}
\begin{figure}[ht]
\includegraphics[width=0.9\columnwidth]{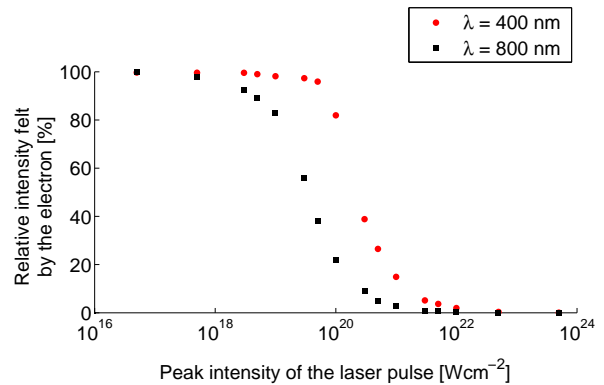}
\caption{The maximum intensity felt by the electron relative the peak intensity of the laser pulse, for different laser intensities. No initial radial displacement of the electron from the propagation axis of the laser pulse. Laser wavelengths are $800\ \mathrm{nm}$ and $400\ \mathrm{nm}$.}
\label{fig:intensity}
\end{figure}

\begin{figure}[ht]
\includegraphics[width=0.9\columnwidth]{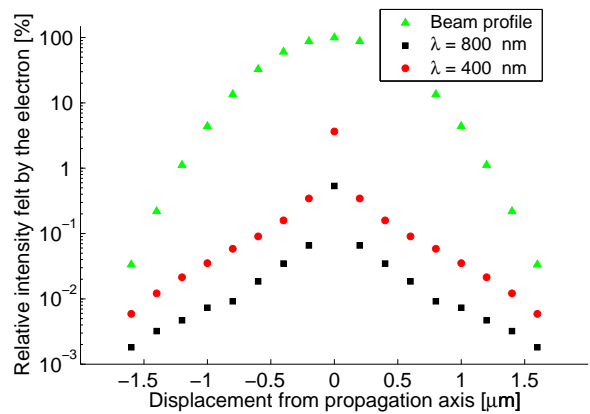}
\caption{The maximum intensity felt by the electron relative the peak intensity of the laser pulse, for different initial radial displacements from the axis of propagation of the laser pulse. The intensity is $5\times10^{21}\ \mathrm{Wcm^{\!-2}}$, and the wavelengths are $800\ \mathrm{nm}$ and $400\ \mathrm{nm}$.}
\label{fig:radialdisplacement}
\end{figure}

To investigate the effect of the ponderomotive force near the focal spot, we have given the laser pulse a Gaussian shape. With the beam parameters given for the Astra Gemini system, simulations show that an electron exposed to such a laser pulse near the interaction region will be forced out of the central beam already during the initial part of the pulse. Fig.\ \ref{fig:intensity} and \ref{fig:radialdisplacement} show the maximum intensity felt by the electron relative to the peak intensity of the laser pulse, for laser pulses with wavelengths $800\ \mathrm{nm}$ and $400\ \mathrm{nm}$. In Fig.\ \ref{fig:intensity}, the intensity of the laser pulse is varied and the electron has no initial radial displacement from the propagation axis of the laser pulse. In Fig.\ \ref{fig:radialdisplacement}, the initial radial displacement is varied for an intensity of $5\times10^{21}\ \mathrm{Wcm^{\!-2}}$. 

Even though the electron blow out leaves a much denser ion plasma, the relative contribution to Compton scattering from ions compared to electrons scales as the square of the mass ratio. We therefore neglect the ion contribution to the competing noise.

\subsection{Compton scattering}
For the intensities considered in this experimental proposal, both linear and non--linear Compton scattering will be present. The non--linear Compton scattering for high intensities have been described both classically \cite{Vachaspati, Esaray} and semiclassicaly \cite{Brown} for an infinite plane wave. The theory predicts generation of a scattered spectra with sharp peaks at each harmonics. This could pose a major problem since 3rd order Compton scattering of the fundamental frequency generates photons with a wavelength that can be confused with the QED signal. However, with a focused laser pulse of this intensity, the broadening of the spectra will be tremendous. This is confirmed in experiment by \cite{Szu-yan}, where the peaks at the harmonics are not distinct.

We are only interested in the number of Compton scattered photons with energies in the same range as the QED scattered photons. Non--linear Compton scattering of 3rd order and higher for 400 nm beam and 4th order and higher for the 800 nm beam are unlikely to yield photons with energies possible to confuse with the QED ones, and can therefore be disregarded. Non--linear Compton effects does not become significant until the dimensionless parameter,
\begin{eqnarray}
	\eta=\frac{e E_{rms}}{\omega m c},
\end{eqnarray}
approaches or exceeds unity \cite{Bamber, Berestetskii}. Here $E_{rms}$ is the root--mean--square value of the electric field amplitude of the laser pulse. Near the interaction region, electrons will be forced out of the central beam already by the leading edge of the pulse, thus experiencing an intensity significantly less than the maximum intensity of the laser pulse. With the given beam parameters and a maximum intensity $I_{\mathrm{max}}$ of $5\times10^{21} \mathrm{Wcm^{\!-2}}$, most electrons will experience an intensity less than $0.5\%$ of $I_{\mathrm{max}}$ (see Fig.\ \ref{fig:radialdisplacement}), giving $\eta\approx1$. For most electrons close to the interaction region, 2nd and 3rd order processes will give a total cross section of the same order of magnitude as that of 1st order Compton scattering \cite{Takahashi, Esaray}. These higher order Compton scattering processes may be the largest source of noise at wavelenths close that of the QED photons. However, we note that we only need a order of magnitude estimate of the noise level. Because of that, and because of the complexity of the problem we are satisfied with estimating the contribution from 1st order compton scattering. 

From now on we will refer to 1st order Compton scattering simply as Compton scattering. We use a hybrid model, where a classical electromagnetic wave accelerates an electron which in its restframe Thomson scatters photons. This opens for inverse Compton scattering, where net energy is transferred from the electron to the photon. We assume that we can filter out and detect only the photons within the wavelength interval $\pm 50\ \mathrm{nm}$ centered at the wavelength of the QED scattered photons, $267\ \mathrm{nm}$. We calculate the scattering cross section for inverse Compton scattering into this wavelength interval, $\sigma_{\Delta}$, in order to give a rough estimate of the noise caused by Compton scattered photons. For simplicity, the incident light is assumed to be monoenergetic with energy $\epsilon$. We further assume that in the rest frame of the electron, the scattering is elastic and isotropic Thomson scattering, which is a decent approximation as long as $\gamma\epsilon\ll mc^{2}$ \cite{lightman}, where $\gamma$ is the relativistic factor. The energy of the scattered photons, $\epsilon_{sc}$, will then be within the range
\begin{eqnarray}
\frac{1-\beta}{1+\beta}<\frac{\epsilon_{sc}}{\epsilon}<\frac{1+\beta}{1-\beta},
\end{eqnarray}
where $\beta=u/c$ and $u$ is the electron velocity. The cross section, $\sigma_\Delta$, for inverse Compton scattering into the energy interval defined by $\epsilon_{low}$ and $\epsilon_{up}$, $\epsilon<\epsilon_{low}<\epsilon_{up}$, can be shown to be (see \cite{lightman})
\begin{equation}\label{eq:compton1}
  \sigma_{\Delta} = 0  
\end{equation}
if ${(1+\beta)}/{(1-\beta)} < {\epsilon_{low}}/\epsilon$,
\begin{equation}\label{eq:compton2}
\sigma_{\Delta} = \frac{\sigma_{T}}{4\epsilon \gamma^{2}\beta^2}\left( \frac{\epsilon}{2}\frac{(1+\beta)^{2}}{1-\beta} - (1+\beta) \epsilon_{low} + (1-\beta) \frac{\epsilon_{low}^{2}}{2\epsilon}\right) 
\end{equation}
if ${\epsilon_{low}}/\epsilon < {(1+\beta)}/{(1-\beta)} < {\epsilon_{up}}/\epsilon$, and 
\begin{equation}\label{eq:compton3}
\sigma_{\Delta} =\frac{\sigma_{T}}{4\epsilon \gamma^{2}\beta^{2}} \left((1+\beta)(\epsilon_{up}-\epsilon_{low}) - (1-\beta)\frac{\epsilon_{up}^{2}-\epsilon_{low}^{2}}{2 \epsilon}\right)
\end{equation}
if ${\epsilon_{up}}/\epsilon < {(1+\beta)}/{(1-\beta)}$. 
Here $\sigma_{T}=6.65\times 10^{-25} \mathrm{cm^{2}}$ is the Thomson scattering cross section.

\begin{center}
\begin{figure}[ht]
\includegraphics[width=0.9\columnwidth]{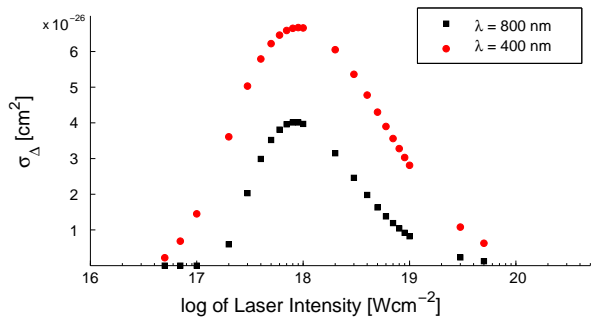}
\caption{The average cross section, $\sigma_{\Delta}$, for inverse Compton scattering of incident light with wavelength $800$ nm and $400$ nm into the wavelength interval $216-316$ nm, as a function of increasing intensity of the incident pulse.}
\label{fig:compton}
\end{figure}
\end{center}

Using the equations of motion for a free electron in an electromagnetic field, $dp^{a}/d\tau=-(e/m)\times F^{a}\!_{b} p^{b}$, the motion of the electron can be simulated and by using (\ref{eq:compton1}) to (\ref{eq:compton3}) the average cross section during a wave period for scattering into a given wavelength interval for a given field intensity can be estimated. Here $p^{a}$ is the electron four--momentum, $\tau$ is the electron eigentime, and $F_{ab}$ is the electromagnetic field tensor. The average scattering cross section for scattering of incident light with wavelength $800$ nm and $400$ nm into the wavelength interval $216-316$ nm for different field intensities is plotted in Fig.\ \ref{fig:compton}. 

An alternative model to the Compton model has also been used to verify the magnitude of the competing scattering events. The model is based on the principle that the laser pulse gives rise to electron oscillations, and for highly relativistic electrons, the parallel component of the acceleration can be neglected compared to the orthogonal one. The radiation is then synchrotron radiation for a particle instantaneously moving in a circular motion of radii $\rho = u^{2}/\dot{u}_{\perp}$ \cite{Jackson}. Simulations have shown that the scattering cross section with this model is roughly the same as that from the Compton model.

\subsection{Collective plasma effects}
In addition to the QED non--linearities, medium non--linearities can also contribute to scattering of photons. For example, in a similar configuration as that considered by us, Ref. \cite{Bernard-exp} has detected photons scattered due to third order susceptibilities in a neutral nitrogen gas. In our case, with field strengths well above the ionization threshold, the pulses will interact with a plasma rather than a gas, and in principle plasma collective non--linearities could compete with the QED effect.  However, with a sufficiently low pressure (of the order $10^{-9}$\,torr), combined with the hole formation caused by the ponderomotive force, only a few electrons will be present within the pulse(s), and none within the central parts of the focused pulse. As a consequence, collective plasma phenomena will not contribute significantly to the scattering made by the electrons.

\subsection{Estimate of Compton scattered photons scattered into the detector}
We now have the tools to make a rough estimate of the number of Compton scattered photons. If the electrons were distributed homogeneously within a pulse of $N_{p}$ photons, the number of scattered photons during a time $\tau$ would be $N_{sc}=\sigma n_{e} c N_{p} \tau$. However, because of the ponderomotive force, most of the photons will not be exposed to any electrons in the region near the focal point, where the intensity is the greatest. We thus introduce a parameter $\kappa$, which is the fraction of photons that do experience electrons. With a maximum intensity of $10^{22}\ \mathrm{Wcm^{\!-2}}$ in the interaction region, we find from Fig.\ \ref{fig:intensity} and \ref{fig:radialdisplacement} that the effect of the ponderomotive force is crucial. We estimate that more than $99\%$ of the beam photons do not encounter any electrons, and we reduce the scattering rate correspondingly, i.e. $\kappa=0.01$.

Only a fraction of the scattered photons will have energies possible to confuse with QED scattered photons, and within that energy range we introduce the subscript $\Delta$ on the cross section $\sigma$. Noting that due to plasma cavitation, the intensity felt by the electrons is significantly less than the peak intensity, we find from Fig.\ \ref{fig:radialdisplacement} the average cross section $\sigma_{\Delta}\approx10^{-26}\ \mathrm{cm^{\!2}}$. From Fig.\ \ref{fig:intensity} and \ref{fig:radialdisplacement} we find that the effect of the ponderomotive force is much more crusial for the $800\ \mathrm{nm}$ pulse than for the $400\ \mathrm{nm}$ one. Furthermore, the average cross section $\sigma_{\Delta}$ is smaller for the $800\ \mathrm{nm}$ photons (see Fig.\ \ref{fig:compton}). Guided by this, we neglect the scattering from the $800\ \mathrm{nm}$ beam compared to that from the two $400\ \mathrm{nm}$ beams. The number of photons in one beam then becomes about $N_{p}\approx6\times 10^{18}$ (40\% of 7.5 J at 400 nm). 

Of the scattered photons, only a fraction, $\delta$, will be scattered into the detector. Electrons accelerated by the laser pulse will oscillate relativisticly in the plane spanned by the polarization vector and the wave vector. Because of the relativistic beaming effect, the Compton scattered photons will be confined close to this plane. If we assume that the detector sees all photons from the QED scattered wave within a cone angle of 30 degrees, this would yield $\delta=0.017$ if all photons were scattered isotropically. However, when considering that electrons are highly relativistic near the interaction region and taking into account the relativistic beaming effect, we estimate $\delta\approx0.01$. The expression for the number of Compton scattered photons per beam reaching the detector is then 
\begin{eqnarray}
	N_{det}= \kappa \delta \sigma_{\Delta} n_{e} c N_{p} \tau.
\end{eqnarray}

We assume that the interaction region is imaged onto the detector in such a way that the detector only sees a part of the interaction region about $10\ \mathrm{\mu m}$ in diameter. This gives us an interaction time of about $\tau\approx3\times10^{-14}\ \mathrm{s}$, during which competing Compton photons can be scattered into the detector. We further assume an unperturbed vacuum around $10^{-9}$ torr, giving $n_{e}\approx 5\times10^{9}\ \mathrm{cm^{\!-3}}$. The number of Compton scattered photons that the detector might confuse with QED scattered photons, from both of the $400\ \mathrm{nm}$ beams, is then about $N_{det}\approx5\times 10^{-5}$, about three orders of magnitude lower than the expected number of QED scattered photons. Here the contribution of higher order Compton scattering has not been included. Even if the non--linear contribution to the noise exceeds the linear one with an order of magnitude, it is still well below the QED signal. Should the noise level in a real experiment still prove to be too severe, it can be further reduced by ionizing the interaction region with a strong laser pulse and extract the electrons with a static electric field shortly before performing the QED experiment.

\section{Conclusions}
In this paper, we have investigated the possibility to detect QED elastic photon--photon scattering through four--wave mixing. In the suggested 3D experimental setup, three colliding laser pulses stimulate emission of a fourth electromagnetic wave with a new frequency and a new direction of propagation. Considering that present day petawatt laser systems only operate at a single frequency (with harmonics available through frequency doubling crystals), it has been shown that a 3D configuration will be able to produce a measurable signal for the Astra Gemini system and similar high power systems. Moreover, by changing the polarization of one of the incoming beams, a unique fingerprint characterizing elastic photon--photon scattering can be obtained. For given data of the laser beams, together with the coupling coefficients, an estimated number of scattered photons was obtained. With the parameters chosen to fit the Astra Gemini system at the Central Laser Facility (Rutherford Appleton Laboratory), the estimated number of scattered photons per shot was $0.07$. Two main obstacles for successful detection were identified, the shot--to--shot reproducibility and the noise from competing scattering events. We have argued that pulse energy fluctuations will be low, and that vibrations are likely to be a small problem due to the inherent structure of the Astra Gemini system. Means of further improvements to reduce vibrations have also been suggested. Due to a non--perfect vacuum in the interaction chamber, there will be competing scattering processes such as Compton scattering and collective plasma four--wave mixing present. Through the use of computer simulations, it was found that the competing effects are suppressed rather effectively due to plasma cavitation caused by the strong laser pulses. With the use of ultra--high vacuum technology, it was found that the noise from competing scattering processes would roughly be about three orders of magnitude lower than the QED signal. Means of further reducing the noise has also been suggested. We conclude that detection of elastic photon--photon scattering will in principle be possible with the Astra Gemini system operational in 2007.

\acknowledgments
J.L.,\ G.B.,\ E.L.\ and M.M.\ would like to thank The Centre for Fundamental Physics, Rutherford Appleton Laboratory. The authors thank S. Hancock for help with illustration. J.L.\ would further like to thank S. Bandyopadhyay, K. Ertel, P. Foster and C. Murphy (Central Laser Facility, Rutherford Appleton Laboratory) for stimulating discussions. 
This research was supported by the Swedish Research Council Contract
No.\ 621-2004-3217.


\appendix

\section{The geometric factor}
The explicit expressions for the general 2D and and the specific 3D geometric factors, $\mathbf{G}_{2d}$ and $\mathbf{G}_{3d}$ respectively, is found below for a Born--Infeld type Lagrangian
\begin{equation}\label{eq:Born-Infeld-Lagrangian}
{\cal L}=\frac{1}{8\pi}\left(E^{2}\!-B^{2}\right)+\frac{\xi}{8\pi}\left[X\left(E^{2}\!-B^{2}\right)^{2}+Y\left(\textbf{E}\cdot\textbf{B}\right)^{2}\right],
\end{equation}
where $X$ and $Y$ are arbitrary coefficients. For the Heisenberg--Euler Lagrangian, $X=1$ and $Y=7$.

\begin{eqnarray}
  &&\!\!\!\!\!\!\!\! \textbf{G}_{2d} = \frac{1}{2}\left\{\ \left[\frac{Y}{4}\cos\gamma_{3}\sin\left(\gamma_{1}+\gamma_{2}\right)-X\sin\gamma_{3}\cos\left(\gamma_{1}+\gamma_{2}\right)\right]\right.\nonumber \\ && \times\left[\sin\left(\phi_{3}-\phi_{4}\right)\cos\phi_{4}-\left(\sin\phi_{3}-\sin\phi_{4}\right)\right]\sin^{2}\left(\frac{\phi_{1}-\phi_{2}}{2}\right)\nonumber \\ &&\ \ \ + \left[\frac{Y}{4}\cos\gamma_{2}\sin\left(\gamma_{1}+\gamma_{3}\right)-X\sin\gamma_{2}\cos\left(\gamma_{1}+\gamma_{3}\right)\right]\nonumber \\ && \times\left[\sin\left(\phi_{2}-\phi_{4}\right)\cos\phi_{4}-\left(\sin\phi_{2}-\sin\phi_{4}\right)\right]\sin^{2}\left(\frac{\phi_{1}-\phi_{3}}{2}\right)\nonumber \\ &&\ \ \ + \left[\frac{Y}{4}\cos\gamma_{1}\sin\left(\gamma_{2}+\gamma_{3}\right)-X\sin\gamma_{1}\cos\left(\gamma_{2}+\gamma_{3}\right)\right]\nonumber \\ &&\!\!\!\!\!\!\!\!  \times\left.\left[\sin\left(\phi_{1}-\phi_{4}\right)\cos\phi_{4}-\left(\sin\phi_{1}-\sin\phi_{4}\right)\right]\sin^{2}\left(\frac{\phi_{2}-\phi_{3}}{2}\right) \right\}\hat{\textbf{x}}\nonumber \\ &&\!\!\!\!+ \frac{1}{2}\left\{\ \left[\frac{Y}{4}\cos\gamma_{3}\sin\left(\gamma_{1}+\gamma_{2}\right)-X\sin\gamma_{3}\cos\left(\gamma_{1}+\gamma_{2}\right)\right]\right.\nonumber \\ &&  \times\left[\sin\left(\phi_{3}-\phi_{4}\right)\sin\phi_{4}+\left(\cos\phi_{3}-\cos\phi_{4}\right)\right]\sin^{2}\left(\frac{\phi_{1}-\phi_{2}}{2}\right)\nonumber \\ &&\ \ \ + \left[\frac{Y}{4}\cos\gamma_{2}\sin\left(\gamma_{1}+\gamma_{3}\right)-X\sin\gamma_{2}\cos\left(\gamma_{1}+\gamma_{3}\right)\right]\nonumber \\ &&  \times\left[\sin\left(\phi_{2}-\phi_{4}\right)\sin\phi_{4}+\left(\cos\phi_{2}-\cos\phi_{4}\right)\right]\sin^{2}\left(\frac{\phi_{1}-\phi_{3}}{2}\right)\nonumber \\ &&\ \ \ + \left[\frac{Y}{4}\cos\gamma_{1}\sin\left(\gamma_{2}+\gamma_{3}\right)-X\sin\gamma_{1}\cos\left(\gamma_{2}+\gamma_{3}\right)\right]\nonumber \\ &&\!\!\!\!\!\!\!\! \times\left.\left[\sin\left(\phi_{1}-\phi_{4}\right)\sin\phi_{4}+\left(\cos\phi_{1}-\cos\phi_{4}\right)\right]\sin^{2}\left(\frac{\phi_{2}-\phi_{3}}{2}\right) \right\}\hat{\textbf{y}}\nonumber \\ &&\!\!\!\!+ \left\{\ \left[\frac{Y}{4}\sin\gamma_{3}\sin\left(\gamma_{1}+\gamma_{2}\right)+X\cos\gamma_{3}\cos\left(\gamma_{1}+\gamma_{2}\right)\right]\right.\nonumber \\ &&\ \ \ \ \times\ \sin^{2}\left(\frac{\phi_{3}-\phi_{4}}{2}\right)\sin^{2}\left(\frac{\phi_{1}-\phi_{2}}{2}\right)\nonumber \\ &&\ + \left[\frac{Y}{4}\sin\gamma_{2}\sin\left(\gamma_{1}+\gamma_{3}\right)+X\cos\gamma_{2}\cos\left(\gamma_{1}+\gamma_{3}\right)\right]\nonumber \\ &&\ \ \ \ \times\ \sin^{2}\left(\frac{\phi_{2}-\phi_{4}}{2}\right)\sin^{2}\left(\frac{\phi_{1}-\phi_{3}}{2}\right)\nonumber \\ &&\ + \left[\frac{Y}{4}\sin\gamma_{1}\sin\left(\gamma_{2}+\gamma_{3}\right)+X\cos\gamma_{1}\cos\left(\gamma_{2}+\gamma_{3}\right)\right]\nonumber \\ &&\ \ \ \ \times\left.\ \sin^{2}\left(\frac{\phi_{1}-\phi_{4}}{2}\right)\sin^{2}\left(\frac{\phi_{2}-\phi_{3}}{2}\right)\ \right\}\hat{\textbf{z}}
  \label{eq:2d}
\end{eqnarray}
\begin{eqnarray}
	\textbf{G}_{3d}\!\!\!\!&=&\!\!\!\!-\frac{2}{9}\left\{\ X\left[\left(\frac{1}{2}\sin\beta_{3}-\cos\beta_{3}\right)\cos\left(\gamma_{1}+\gamma_{2}\right)\right.\right.\nonumber \\&&\ \ \ \ \ \ +\left(\frac{1}{8}\sin\gamma_{2}-\frac{1}{4}\cos\gamma_{2}\right)\cos\left(\gamma_{1}+\beta_{3}\right)\nonumber \\&&\ \ \ \ \ \ +\!\left.\left(\frac{1}{4}\sin\gamma_{1}+\frac{1}{8}\cos\gamma_{1}\right)\sin\left(\gamma_{2}+\beta_{3}\right)\right]\nonumber \\&&\ \ \ +\frac{Y}{4}\left[\left(-\sin\beta_{3}-\frac{1}{2}\cos\beta_{3}\right)\sin\left(\gamma_{1}+\gamma_{2}\right)\right.\nonumber \\&&\ \ \ \ \ \ \ \  +\left(-\frac{1}{8}\cos\gamma_{2}-\frac{1}{4}\sin\gamma_{2}\right)\sin\left(\gamma_{1}+\beta_{3}\right)\nonumber \\&&\ \ \ \ \ \ \ \  +\!\!\left.\left.\left(\frac{1}{4}\cos\gamma_{1}-\frac{1}{8}\sin\gamma_{1}\right)\cos\left(\gamma_{2}+\beta_{3}\right)\right]\ \right\}\hat{\textbf{x}}\nonumber \\&&\!\!\!\! -\frac{2}{9}\left\{\ X\left[\left(\frac{1}{2}\cos\beta_{3}-\sin\beta_{3}\right)\cos\left(\gamma_{1}+\gamma_{2}\right)\right.\right.\nonumber \\&&\ \ \ \ \ \ +\left(\frac{1}{8}\cos\gamma_{2}-\frac{1}{4}\sin\gamma_{2}\right)\cos\left(\gamma_{1}+\beta_{3}\right)\nonumber \\&&\ \ \ \ \ \  +\!\left.\left(-\frac{1}{4}\cos\gamma_{1}-\frac{1}{8}\sin\gamma_{1}\right)\sin\left(\gamma_{2}+\beta_{3}\right)\right]\nonumber \\&&\ \ \  +\frac{Y}{4}\left[\left(\cos\beta_{3}+\frac{1}{2}\sin\beta_{3}\right)\sin\left(\gamma_{1}+\gamma_{2}\right)\right.\nonumber \\&&\ \ \ \ \ \ \ \  +\left(\frac{1}{8}\sin\gamma_{2}+\frac{1}{4}\cos\gamma_{2}\right)\sin\left(\gamma_{1}+\beta_{3}\right)\nonumber \\&&\ \ \ \ \ \ \ \  +\!\!\left.\left.\left(\frac{1}{4}\sin\gamma_{1}-\frac{1}{8}\cos\gamma_{1}\right)\cos\left(\gamma_{2}+\beta_{3}\right)\right]\ \right\}\hat{\textbf{y}}\nonumber \\&&\!\!\!\! +\frac{4}{9}\left\{\ X\left[\frac{1}{2}\left(\cos\beta_{3}+\sin\beta_{3}\right)\cos\left(\gamma_{1}+\gamma_{2}\right)\right.\right.\nonumber \\&&\ \ \ \ \ \ +\frac{1}{8}\left(\sin\gamma_{2}+\cos\gamma_{2}\right)\cos\left(\gamma_{1}+\beta_{3}\right)\nonumber \\ &&\ \ \ \ \ \ \!\left.+\frac{1}{8}\left(\cos\gamma_{1}-\sin\gamma_{1}\right)\sin\left(\gamma_{2}+\beta_{3}\right)\right]\nonumber \\&&\ \ \ +\frac{Y}{4}\left[\frac{1}{2}\left(\sin\beta_{3}-\cos\beta_{3}\right)\sin\left(\gamma_{1}+\gamma_{2}\right)\right.\nonumber \\&&\ \ \ \ \ \ \ \ +\frac{1}{8}\left(\sin\gamma_{2}-\cos\gamma_{2}\right)\sin\left(\gamma_{1}+\beta_{3}\right)\nonumber \\&& \!\!\!\! + \left.\left.\frac{1}{8}\left(-\cos\gamma_{1}-\sin\gamma_{1}\right)\cos\left(\gamma_{2}+\beta_{3}\right)\right] \right\}\hat{\textbf{z}}
	\label{eq:3d}
\end{eqnarray}

\end{document}